# Fingerprinting of Compound-Specific Chlorine Isotopologue Distribution of Organochlorines by GC-HRMS

**Caiming Tang[1,2,]* and Jianhua Tan[3]**

[1] *State Key Laboratory of Organic Geochemistry, Guangzhou Institute of Geochemistry, Chinese Academy of Sciences, Guangzhou 510640, China*

[2] *University of Chinese Academy of Sciences, Beijing 100049, China*

[3] *Guangzhou Quality Supervision and Testing Institute, Guangzhou, 510110, China*

*Corresponding Author.

Tel: +86-020-85291489; Fax: +86-020-85290009; E-mail: CaimingTang@gig.ac.cn. (C. Tang).

**ABSTRACT**

This study developed an analytical method to measure compound-specific chlorine isotopologue distribution of organochlorines for source identification and apportionment. Complete chlorine isotopologues of individual model organochlorines were detected by gas chromatography-double focus magnetic-sector high resolution mass spectrometry (GC-DFS-HRMS). The measured relative abundances ($RA_{mea}$), simulated relative abundances ($RA_{sim}$), and relative variations between $RA_{mea}$ and $RA_{sim}$ ($\Delta RA$) were obtained on basis of the detected MS signal intensities of individual isotopologues. The method has been partially validated in terms of precision, injection-amount dependency and temporal drifts. The standard deviations (SDs) of $RA_{mea}$ of all istotopologues of perchlorethylene (PCE) and trichloroethylene (TCE) from different manufacturers were 0.02‰-0.69‰. The SDs of $\Delta RA$ of the first three isotopologues of PCE and TCE from different manufacturers were 0.26‰-1.55‰. The $\Delta RA$ and $\Delta RA$ patterns of the standards of PCE and TCE from different manufacturers were able to be differentiated with statistical significance. The $\Delta RA$ showed no observable injection-amount dependency and no evident temporal drift with the injection times and analysis batches. The method has been successfully applied to an analogous case of source identification and apportionment for two trichlorodiphenyls that exhibited significant chlorine isotope fractionation on the GC-HRMS system. The results demonstrate that the $\Delta RA$ in conjunction with isotope ratios can be applied to source identification and present high-validity identification outcomes. Moreover, the $RA_{mea}$ can be used to implement source apportionment for organochlorines from more sources with more reliable outcomes compared with the methods using isotope ratios only. This method opens a new way to perform fingerprinting analysis of compound-specific chlorine isotopologue distribution of organochlorines, and will be a promising high-performance approach in source delineation and apportionment for chlorinated organic compounds.

## INTRODUCTION

Chlorinated organic compounds have raised lots of public and scientific concerns due to their widespread distribution, occurrence and applications.[1] A large number of organochlorines are ubiquitous and notorious environmental pollutants and toxic substances, including anthropogenic and naturally occurring compounds.[2-7] Exploring sources and transformation processes of organochlorine pollutants in the environment is of high significance to evaluate their potential influences on human beings and eco-environmental systems.[1,8] Source identification and apportionment of contaminants are crucial and difficult tasks in environmental study fields, which involve identifying pollution sources and quantifying attribution proportions from individual sources.[9-11] Up to now, many methods have been developed and applied to source apportionment for pollutants,[12] e.g., chemical mass balance,[13,14] multivariate statistics,[15] hybrid approaches,[16] molecular marker method,[17] and isotope techniques.[1,18] Isotope-signature approaches, especially compound-specific isotope analysis (CSIA), are useful and practical in source apportionment for environmental pollutants.[1] The recently developed dual CSIA methods for two elements are even more powerful in pollution source apportionment compared with the CSIA methods for single element.[18]

CSIA of chlorine (CSIA-Cl) has been applied to source delineation and apportionment of chlorinated organic pollutants, showing promising capability in these tasks.[19,20] CSIA-Cl can be conducted with several instruments, such as preparative gas chromatography (GC) offline isotope ratio mass spectrometry (IRMS),[21-26] GC coupled with IRMS (GC-IRMS),[23] GC-inductively coupled plasma MS (GC-ICP-MS),[27] GC-high-temperature conversion (HTC)-MS/IRMS (GC-HTC-MS/IRMS),[28] GC-quadrupole MS (GC-qMS),[29-34] GC-hybrid quadrupole time of flight MS (GC-QTOF-MS)[35] and GC-double focus magnetic-sector high resolution MS (GC-DFS-HRMS).[36] Preparative GC offline IRMS for CSIA-Cl requires laborious sample preparation and cumbersome conversion of target organochlorines into simple molecules containing only one Cl atom (e.g., CsCl and $CH_3Cl$) on IRMS,[37] and the required injection amount is relatively high. GC-IRMS and GC-ICP-MS can be used for CSIA-Cl without conversion of analytes, but they can only be applicable for a narrow spectrum of organochlorines with relatively high limits of detection (LODs).[30,33] The CSIA-Cl using GC-HTC-IRMS/MS presented relatively high LODs also and required high-temperature conversion of target organochlorines into HCl.[28] CSIA-Cl methods using GC-qMS with electron ionization (EI) have been developed during the last decade, showing comparable precision and accuracy with the methods using conventional GC-IRMS in some cases and obvious advantages including high sensitivity as well as convenient accessibility and operation.[29,34] So far, the reported CSIA-Cl methods using GC-qMS and GC-QTOF-MS applied the isotope-ratio calculation schemes using pair(s) of neighboring chlorine isotopologues of molecular and/or product ions of analytes, which were based on the prerequisite hypothesis that the detected abundances of all isotopologues of a certain ion conform to binomial distribution.[29,31,37] However, our recent studies in both experiment and principle have proved that chlorine isotopologues of detected ions on EI-MS by no means comply with binomial distribution.[38-40] Thus, the reported CSIA-Cl methods using GC-qMS and GC-QTOF-MS need reconsideration and reevaluation. Recently, we developed a CSIA-Cl method by using GC-DFS-HRMS in association with a complete-isotopologue scheme of isotope-ratio calculation using molecular ions only to avoid the disadvantages caused by non-binomial distributions of chlorine isotopologues of the detected ions of target organochlorines.[40] In addition, the complete-isotopologue scheme is suitable for GC-DFS-HRMS rather than GC-qMS, because GC-DFS-HRMS can provide significantly higher sensitivity and selectivity for organochlorines than GC-qMS, particularly for polychlorinated organic compounds with more than two Cl atoms.

At present, only bulk (average) isotope compositions of organochlorine pollutants are analyzed and used for source apportionment. Nevertheless, it is possible that the bulk chlorine isotope ratios of an organochlorine from different sources are equal while the chlorine isotopologue distributions are different. Therefore, the bulk compound-specific isotope ratios may mask some important information about the analytes in different samples in terms of isotopologue distribution, and thus lead to misjudgments in source identification and apportionment. Isotopologue distributions are theoretically more compound-specific and capable of providing more informative clues for source apportionment. GC-DFS-HRMS has been proved to enable the detection of complete chlorine isotopologues of molecular ions with Cl atoms up to six, with reasonable precision at a feasible injection amount around 0.5 ng.[36] As a result, it is anticipatable that GC-DFS-HRMS can implement the analysis of compound-specific chlorine isotopologue distribution of organochlorines with rational and acceptable precision, accuracy and sensitivity.

In this study, we applied GC-DFS-HRMS to developing a method for fingerprinting and determination of compound-specific chlorine isotopologue distribution and isotope ratios of organochlorines for source identification and apportionment. The method has been partially validated in terms of precision, sensitivity, injection-amount dependency and temporal drifts, and successfully applied to an analogous study of source identification and apportionment of two model organochlorines exhibiting chlorine isotope fractionation on the GC-HRMS. This method opens a new door to characterize and analyze compound-specific chlorine isotopologue distribution of organochlorines, and will become a promising tool in source apportionment for chlorinated organic compounds.

# EXPERIMENTAL SECTION

**Chemicals and materials.**

Chromatographic-grade standards of perchlorethylene (PCE, 99.0%) and trichloroethylene (TCE, 99.5%) were purchased from Dr. Ehrenstorfer (Augsburg, Germany, manufacturer-1), and the analytical-grade PCE and TCE were obtained from Tianjin Fuyu Chemical Co. Ltd. (Tianjin, China, manufacturer-2). Standard solutions (10.0 μg/mL in isooctane) of polychlorinated biphenyls (PCB-18 and PCB-28) were bought from Accustandard Inc. (New Haven, CT, USA). Full names, abbreviations, structures, and other relevant information of the chemicals are documented in the Supporting Information (Table S-1). Isooctane and n-hexane were of chromatographic grade and obtained from CNW Technologies GmbH (Düsseldorf, Germany) and Merck Corp. (Darmstadt, Germany), respectively.

The standards of PCE and TCE were accurately weighed and subsequently dissolved with n-hexane to obtain stock solutions at 1.0 mg/mL. The stock solutions of PCE and TCE and the purchased standard solutions of PCB-18 and PCB-28 were further serially diluted with n-hexane or isooctane to prepare working solutions at different concentration levels (e.g., 1.0 and 0.1 μg/mL). All the standard solutions were kept at -20 ºC condition before use.

**Instrumental analysis.**

The GC-HRMS system consisted of dual gas chromatographers (Trace-GC-Ultra) coupled with a double focus magnetic-sector HRMS and a Triplus auto-sampler (GC-DFS-HRMS, Thermo-Fisher Scientific, Bremen, Germany). A capillary GC column (DB-5MS, 60 m × 0.25 mm, 0.25 μm thickness, J&W Scientific, USA) was utilized, and helium was used as the carrier gas with a constant flow rate at 1.0 mL/min. The oven temperature program for analyzing PCE and TCE was: initially held at 40 °C for 2 min, ramped at 2 °C/min to 65°C, then ramped to 300 ºC at 40 ºC/min and held for 1 min. The temperature program for analysis of PCB-18 and PCB-28 is detailed in Table S-1. The GC inlet and transfer line were set at 260 ºC and 280 ºC, respectively.

The working parameters and conditions of the HRMS are documented as follows: electron ionization source in positive mode (EI+) was used; EI energy was 45 eV or 70 eV; ionization source was kept at 250 ºC; filament current of the EI source was 0.8 mA; multiple ion detection (MID) mode was applied for data acquisition; dwell time of each isotopologue ion was around 20 ms; mass resolution was $\geq$ 10000 (5% peak-valley definition) and the HRMS detection accuracy was within ± 0.001 u. The MID began at 7.6 min, and the detecting time segments were 7.6-11.5 min and 11.5-15 min for TCE and PCE, respectively. The MID cycle for PCE was of 140 ms, and that for TCE was of 120 ms. The HRMS was calibrated with perfluorotributylamine during MID operation.

Chemical structures of the investigated compounds were drawn with ChemDraw (Ultra 7.0, Cambridgesoft), and the exact masses of the molecular isotopologues were calculated with mass accuracy of 0.00001 u. Only chlorine isotopologues were considered. For a compound with $n$ Cl atoms, all the isotopologues ($n$ + 1) were selected. The mass-to-charge ratio ($m/z$) of each isotopologue ion was calculated by subtracting the mass of an electron from the mass of the corresponding molecular isotopologue. The $m/z$ values were imported into the MID module for instrumental detection. The detailed data relevant to isotopologues of the investigated compounds, e.g., retention time, isotopologue chemical formulas, exact masses and exact $m/z$ values are listed in Table S-2.

The GC-qMS system was comprised of an Agilent 7890 GC coupled with a 5975 MS (Agilent Technologies, Palo Alto, CA, USA). The column used in the GC-qMS was with the same specification of that used in the in the GC-HRMS system. Detailed descriptions of the temperature programs are provided in Table S-1. The working parameters of the qMS are provided as follows: positive EI+ was used; EI energy was set at 45 eV or 70 eV; the source temperature was maintained at 230 ºC; selective ion monitoring (SIM) mode was applied to data acquisition; dwell time of every single isotopologue ion was 30 ms; mass resolution was set at high level (0.2 u). The details including *m/z* values of isotopologue ions for SIM are provided in Table S-2.

The working solutions were directly injected onto the GC-HRMS or GC-qMS. The working solutions (1.0 and 0.1 µg/mL) of PCE and TCE standards stemming from different producers were detected by the GC-HRMS with a bracketing injection mode as proposed in a previous study,[30] and 5-6 injection replicates were conducted for each working solution in each analysis batch. The PCE and TCE standards from manufacturer-1 were analyzed by GC-qMS also, with the concentration of 100.0 µg/mL. PCB-18 and PCB-28 were merely analyzed by GC-HRMS, with the concentration of 1.0 µg/mL.

**Data processing.**

Chlorine isotope ratio (IR) was calculated by

$$IR = \frac{\sum_{i=0}^{n} i \times I_i}{\sum_{i=0}^{n} (n-i) \times I_i} \qquad (3)$$

where *n* is the number of Cl atoms of an organochlorine; *i* represents the number of $^{37}$Cl atoms in an isotopologue ion; $I_i$ denotes the MS signal intensity of the isotopologue ion *i*. The measured relative abundance of a chlorine isotopologue ($RA_{mea}$) of the organochlorine was calculated as:

$$RA_{mea} = \frac{I_i}{\sum_{i=0}^{n} I_i} \qquad (4)$$

We hypothesized that the theoretical relative abundances of chlorine isotopologues of each investigated polychlorinated compound conformed to binomial distribution. Therefore, in principle, the theoretically simulated relative abundance of a chlorine isotopologue ($RA_{sim}$) of the organochlorine can be calculated with

$$RA_{sim} = \binom{n}{i} \left(\frac{1}{1+IR}\right)^{n-i} \left(\frac{IR}{1+IR}\right)^{i} \quad (5)$$

Accordingly, the relative-abundance variation of a chlorine isotopologue ($\Delta RA$) can be expressed as follows:

$$\Delta RA = \left(\frac{RA_{mea}}{RA_{sim}} - 1\right) \times 1000‰ \quad (6)$$

All the measured isotope ratios and relative abundances of chlorine isotopologues in this study were raw values without being calibrated to Standard Mean Ocean Chlorine (SMOC) scale due to unavailability of the external isotopic standards with known chlorine isotope composition and identical structures to the investigated compounds.

The average MS signal intensity of each isotopologue ion derived from individual whole chromatographic peak was used for calculating isotope ratios, $RA_{mea}$ and $RA_{sim}$. Background subtraction was carried out before exporting MS signal intensity by subtracting intensities of the baseline regions neighboring both ends of the corresponding chromatographic peak. Data from 5-6 replicated injections were applied to evaluating the average isotope ratios, $RA_{mea}$ and $RA_{sim}$, along with their standard deviations (SD, 1σ).

## RESULTS AND DISCUSSION

**Method performances.**

*Precision.* As documented in Table 1, the SDs of $RA_{mea}$ and $RA_{sim}$ of the five chlorine istotopologues (IST-1 to IST-5) of PCE from manufacturer-1 were 0.02‰-0.69‰ and 0.01‰-0.39‰, respectively. The SDs of $RA_{mea}$ and $RA_{sim}$ of PCE from manufacturer-2 were 0.02‰-0.39‰ and 0.01‰-0.29‰, respectively. For TCE from the two manufacturers, the precisions of $RA_{mea}$ and $RA_{sim}$ of corresponding individual isotopologues were generally comparable, with SDs in the range of 0.04‰-0.57‰.

Since ΔRA were the essential data in this study, the precision in analysis of ΔRA was thus vital. As show in Table 1, the SDs of ΔRA of the first three isotopologues (IST-1 to IST-3) of PCE and TCE from manufacturer-1 were 0.26‰-1.55‰, and those of PCE and TCE from manufacturer-2 were 0.26‰-0.84‰. For the fourth isotopologues (IST-4) of PCE and TCE, their SDs of ΔRA were relatively higher than those of the first three isotopologues, with the range of 2.20‰-4.89‰. The fifth isotopologue (IST-5) of PCE showed the highest SDs of ΔRA, with the ranges of 7.61‰-16.91‰ and 5.94‰-10.37‰ for the standards from manufacturer-1 and manufacturer-2, respectively.

As illustrated in Figure 1, the ΔRA of IST-1 to IST-3 of PCE from the different manufacturers were able to be confidently differentiated, unequivocally demonstrating the different isotopologue distributions and sources of the PCE standards. Whereas the precision of ΔRA for IST-4 and IST-5 of PCE could not be competent in distinguishing the corresponding ΔRA of the standards from the two manufacturers. On the other hand, the ΔRA of all the isotopologues (IST-1 to IST-4) of TCE from different manufacturers were evidently distinguishable, showing the different distributions of chlorine isotopologues of the two standards. Although the ΔRA of some higher-mass and lower-intensity isotopologues of PCE from different sources could not be confidently distinguished due to the insufficient precision, the ΔRA patterns of the two standards could be identified as definitely different (Figure 1). For confirmation, the chlorine isotope ratios of PCE and TCE were calculated also. The isotope ratios of PCE and TCE from different manufacturers were also clearly distinguishable, indicating different chlorine isotope compositions (Figure 2).

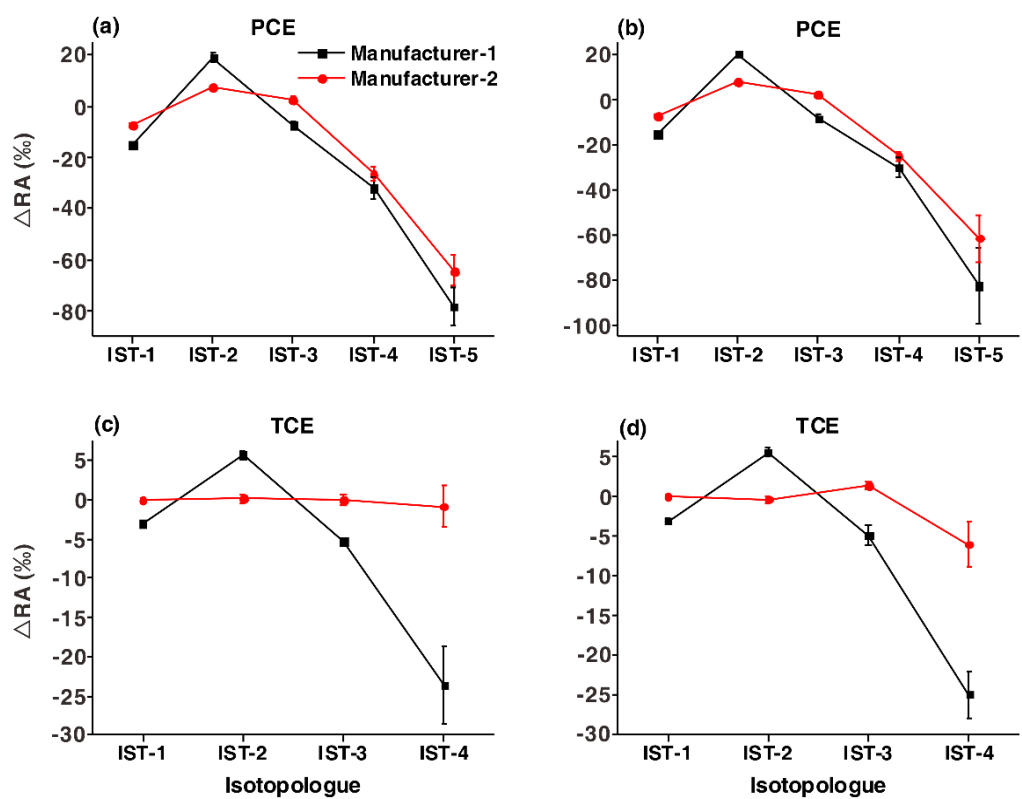

**Figure 1**. Relative-abundance variations of chlorine isotopologues (ΔRA) of perchlorethylene (PCE) and trichloroethylene (TCE) from different manufacturers in different analysis batches with the detection of GC-HRMS. **(a)**: ΔRA of PCE in batch-I, **(b)**: ΔRA of PCE in batch-II, **(c)**: ΔRA of TCE in batch-I, **(d)**: ΔRA of TCE in batch-II. IST: isotopologue; IST-$x$: a chlorine isotopologue with $x - 1$ $^{37}$Cl atom(s); the standard solutions were at 1.0 μg/mL; the injection replicates of batch-I and batch-II were 5 and 6, respectively; error bars denote the standard deviations (SDs).

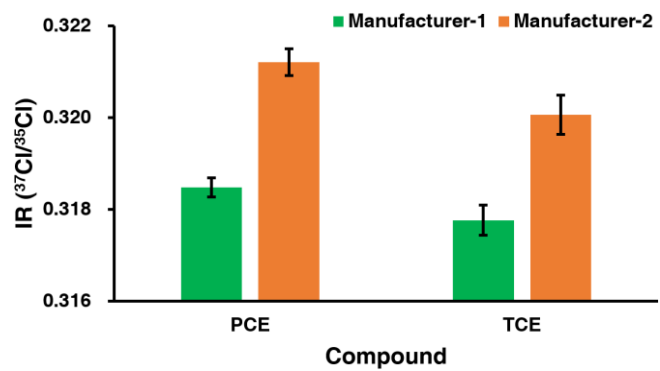

**Figure 2**. Measured isotope ratios of PCE and TCE from different manufacturers with the detection of GC-HRMS. The standard solutions were at 1.0 μg/mL; IR: isotope ratio ($^{37}Cl/^{35}Cl$); the injection replicates were 6.

*Injection-amount dependency.* The ΔRA of PCE and TCE at the concentrations of 1.0 and 0.1 μg/mL (1.0 and 0.1 ng on column) were analyzed with GC-HRMS. As shown in Figure 3, the ΔRA patterns of PCE and TCE at different concentrations were fairly consistent, particularly for TCE. This indicates that the ΔRA might be independent on injection concentrations/amounts, even though the precision of ΔRA became lower when the injection amount was lower.

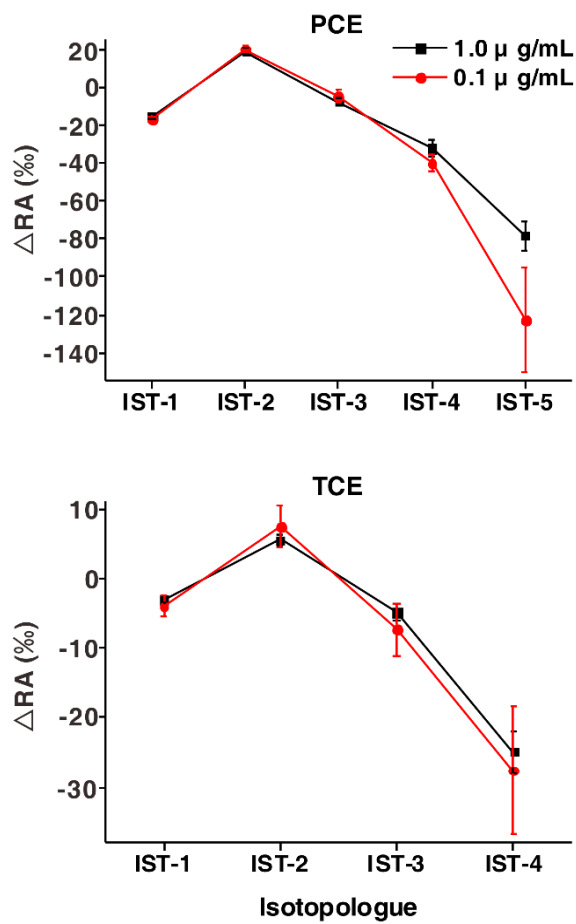

**Figure 3**. ΔRA of PCE and TCE at different concentrations measured with GC-HRMS.

**Measured ΔRA on different MS with different EI energies.**

As revealed in our previous studies, the isotope fractionation behaviors taking place on one MS with different EI energies could be inconsistent, and also different on different types of MS.[41,42] We thus anticipated that the ΔRA patterns of organochlorines might be different on an MS with different EI energies or on different types of MS. As illustrated in Figure 4a, b, PCE showed obviously different ΔRA patterns on either HRMS or qMS, at the EI energies of 45 eV and 70 eV. The difference of ΔRA patterns of PCE on HRMS at different EI energies was more significant than that on qMS with different EI energies. Figure 4b clearly illustrates the difference of ΔRA patterns of PCE on qMS with the EI energies of 45 eV and 70 eV. TCE showed significantly different ΔRA patterns on different types of MS, while its ΔRA patterns on respective MS with different EI energies were very consistent. The results also indicated that the differences of ΔRA patterns caused by different types of MS were much larger than those caused by different EI energies. This finding might be attributable to the different types of multiple dechlorination reactions (concerted or stepwise) occurring in the two types of MS.[42]

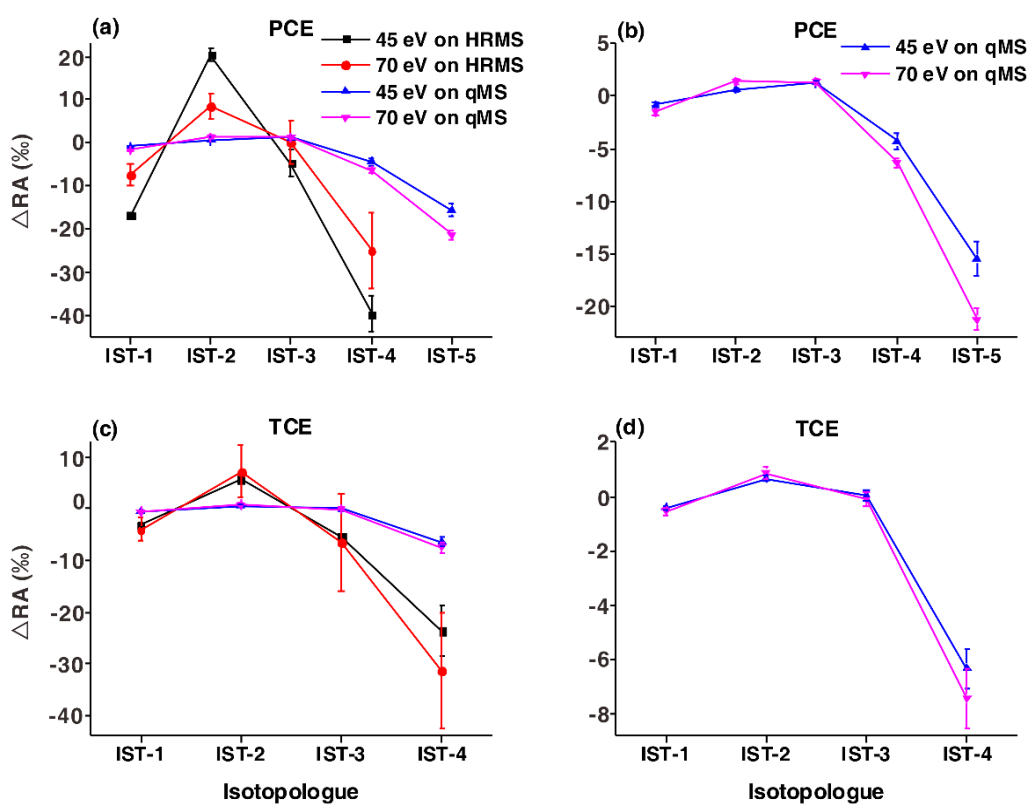

**Figure 4**. ΔRA of PCE and TCE measured with GC-HRMS and GC-qMS with different electron ionization (EI) energies. **(a)**: ΔRA of PCE measured with GC-HRMS and GC-qMS, **(b)**: ΔRA of PCE measured with GC-qMS, **(c)**: ΔRA of TCE measured with GC-HRMS and GC-qMS, **(d)**: ΔRA of TCE measured with GC-qMS. The standard solutions were at 0.1 μg/mL for GC-HRMS and 100.0 μg/mL for GC-qMS, respectively; the injection replicates of batch-I and batch-II were 5 and 6, respectively.

**Temporal drifts of ΔRA with injection sequences and analysis batches.**

Our previous study demonstrated that the isotope ratios of the investigated organochlorines were consistent with the injection sequence within an analysis batch, whereas TCE showed temporal drifts of isotope ratios between two independent analysis batches. In this study, the first five injections (1-5) and the rest six injections (6-11) belonged to batch-I and batch-II, respectively. As shown in Figure 5, the ΔRA of all the isotopologues of PCE and TCE from all sources did not show a systematic temporal drift with the injection sequence. For PCE, the first three isotopologues (IST-1 to IST-3) showed more constant ΔRA than the last two isotopologues (IST-4 to IST-5), and IST-5 presented the largest ΔRA variations among different injections. Similarly, the ΔRA of the first three isotopologues of TCE were much more constant than those of the last isotopologue (IST-4). The ΔRA of the five isotopologues of PCE from the two manufacturers showed the following order: IST-2 > IST-3 > IST-1 > IST-4 > IST-5. The ΔRA of all the isotopologues of TCE from manufacturer-1 followed the order as: IST-2 > IST-1 > IST-3 > IST-4. Nevertheless, the TCE standard from manufacturer-2 showed indistinguishable ΔRA from IST-1 to IST-4.

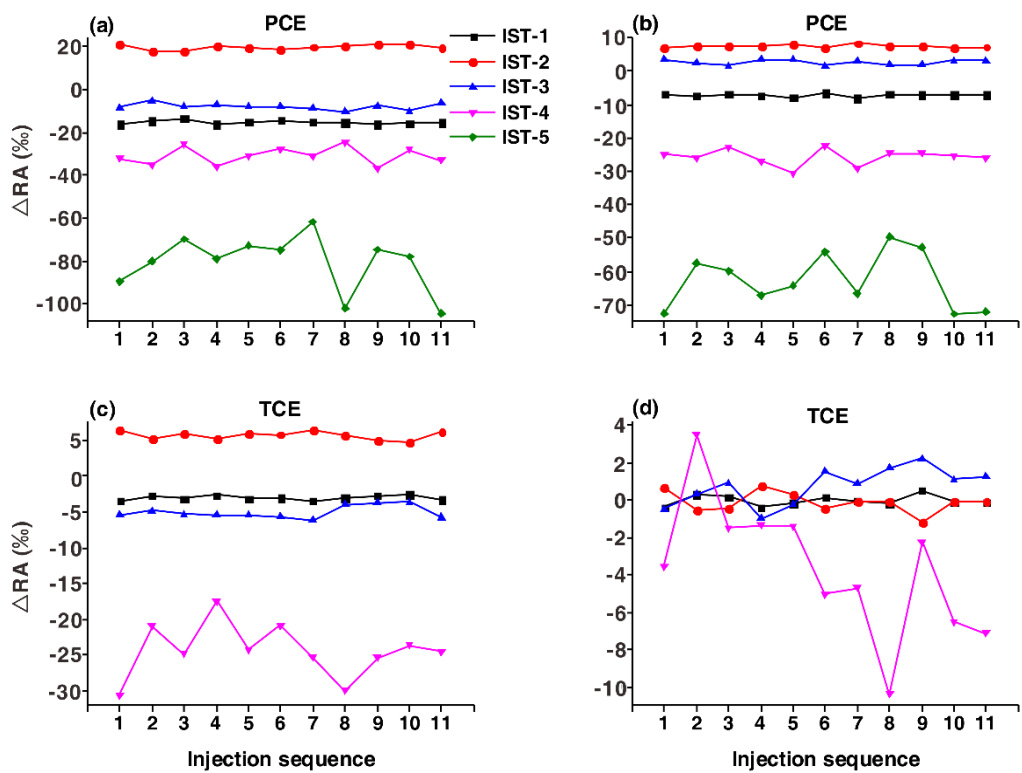

**Figure 5**. Temporal drifts of ΔRA with injection times of PCE and TCE with the detection of GC-HRMS. **(a)**: PCE from manufacturer-1, **(b)**: PCE from manufacturer-2, **(c)**: TCE from manufacturer-1, **(d)**: TCE from manufacturer-2. The standard solutions were at 1.0 μg/mL.

As shown in Figure 6, the ΔRA patterns of PCE and TCE derived from different analysis batches are consistent, especially for the two PCE standards and the TCE standard from manufacturer-1. With respect to the last three isotopologues of the TCE standard from manufacturer-2, the ΔRA of individual isotopologues in different batches showed no statistically significant difference (Figure 6d), even though the corresponding ΔRA were not in accordance very well with each other compared with the ΔRA of other standards.

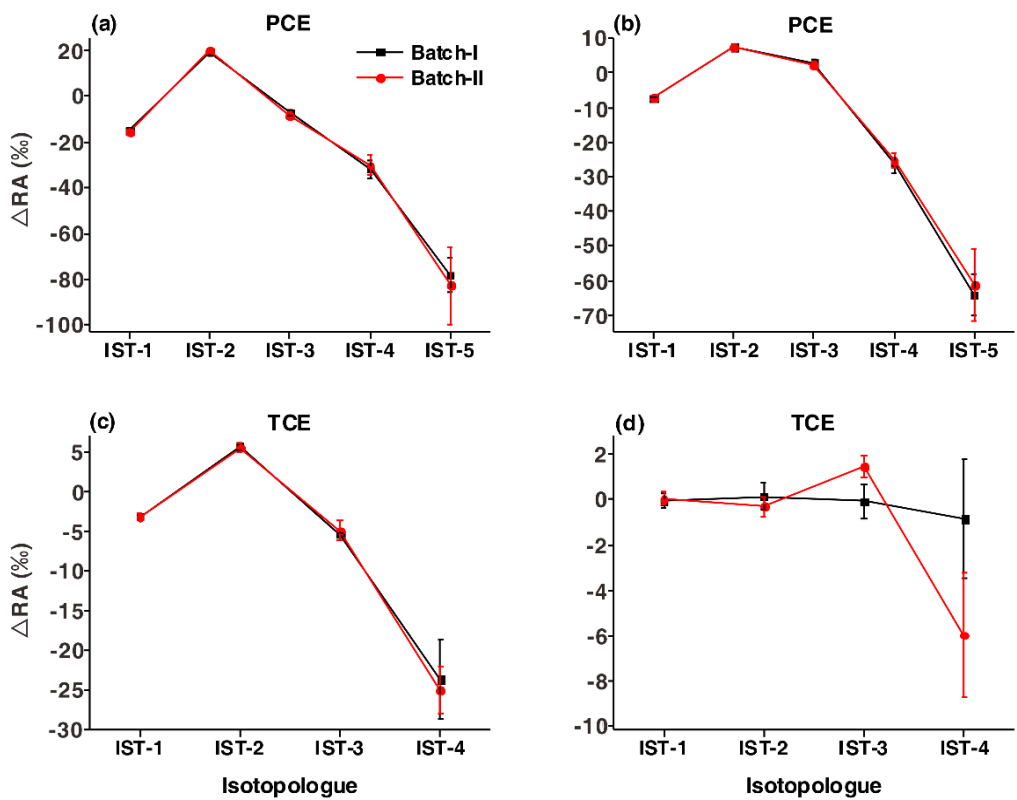

**Figure 6**. ΔRA of PCE and TCE in different analysis batches with the detection of GC-HRMS. **(a)**: PCE from manufacturer-1, **(b)**: PCE from manufacturer-2, **(c)**: TCE from manufacturer-1, **(d)**: TCE from manufacturer-2. The interval between the two batches was one day; the standard solutions were at 1.0 μg/mL.

**Variations of ΔRA during physical and chemical changes of PCB-18 and PCB-28.**

Since PCE and TCE could not present chlorine isotope fractionation on the GC column as found in our previous study,[40] we used the data relevant to chlorine isotope fractionations of PCB-18 and PCB-28 on GC and in EI-HRMS to investigate the ΔRA variations of organochlorines during physical and chemical changes. As Figure 7a, b show, the ΔRA patterns of PCB-18 and PCB-28 derived from different retention-time segments and the overall retention-time ranges were highly similar, indicating consistent ΔRA in different retention-time segments. On the other hand, PCB-18 and PCB-28 exhibited significant inverse chlorine isotope fractionation on the GC column, with the isotope fractionation extents ($\Lambda^{37}Cl$) of 73.1‰ and 46.5‰ between the first retention-time segment (T1) and the last segment (T5), respectively. GC separation is a physical change involving processes such as adsorption, desorption, dissolution, evaporation, diffusion and condensation, during which isotope fractionation may take place. Isotope fractionation occurring in physical changes manifests as variations of isotope ratios in different phases or parts of compounds, in other words, isotope ratios vary when physical changes occur. However, the results in this study revealed that the ΔRA of chlorine isotopologues of organochlorines were insusceptible to the physical change of GC separation, even though evident isotope fractionation was observed during the separation (Figure 7a-d). Accordingly, we conclude that the ΔRA and ΔRA patterns may not change during physical changes. This inference may be very useful in source apportionment of organochlorines, because it can reduce the misjudgments in source identification in some cases. For instance, if two groups of an organochlorine are from one source but have experienced different physical changes that can lead to isotope fractionation and isotope-ratio deference, then applying the isotope ratios to source identification may result in a mistake, i.e., they are determined as coming from different sources. Therefore, application of ΔRA patterns in association with isotope ratios is a promising approach to reduce misjudgments in source identification and apportionment for organochlorines.

As Figure 4 and Figure 7e, f show, the ΔRA patterns of PCE and TCE measured with different types of MS were evidently different, and those of PCE, PCB-18 and PCB-28 obtained by the same MS at different EI energies were also statistically different. Dechlorination reactions of organochlorines in EI-MS are chemical changes. Different types of MS and different EI energies may lead to different types and/or extents of dechlorination reactions,[41,42] which could give rise to different ΔRA patterns and isotope ratios of organochlorines (Figure 2 and Figure 7e-h). The same dechlorination-reaction type and extent may lead to identical ΔRA patterns. Therefore, if two groups of an organochlorine from one source have identical ΔRA patterns which are different from the initial, it can be deduced that they have experienced the same dechlorination process. ΔRA patterns can also be used to distinguish the isotope-ratio differences caused by physical changes from those triggered by chemical changes, because the ΔRA patterns during physical processes are consistent and those after chemical reaction processes are different. With ΔRA patterns and isotope ratios, it is possible to deduce the chemical reaction processes experienced by organochlorines.

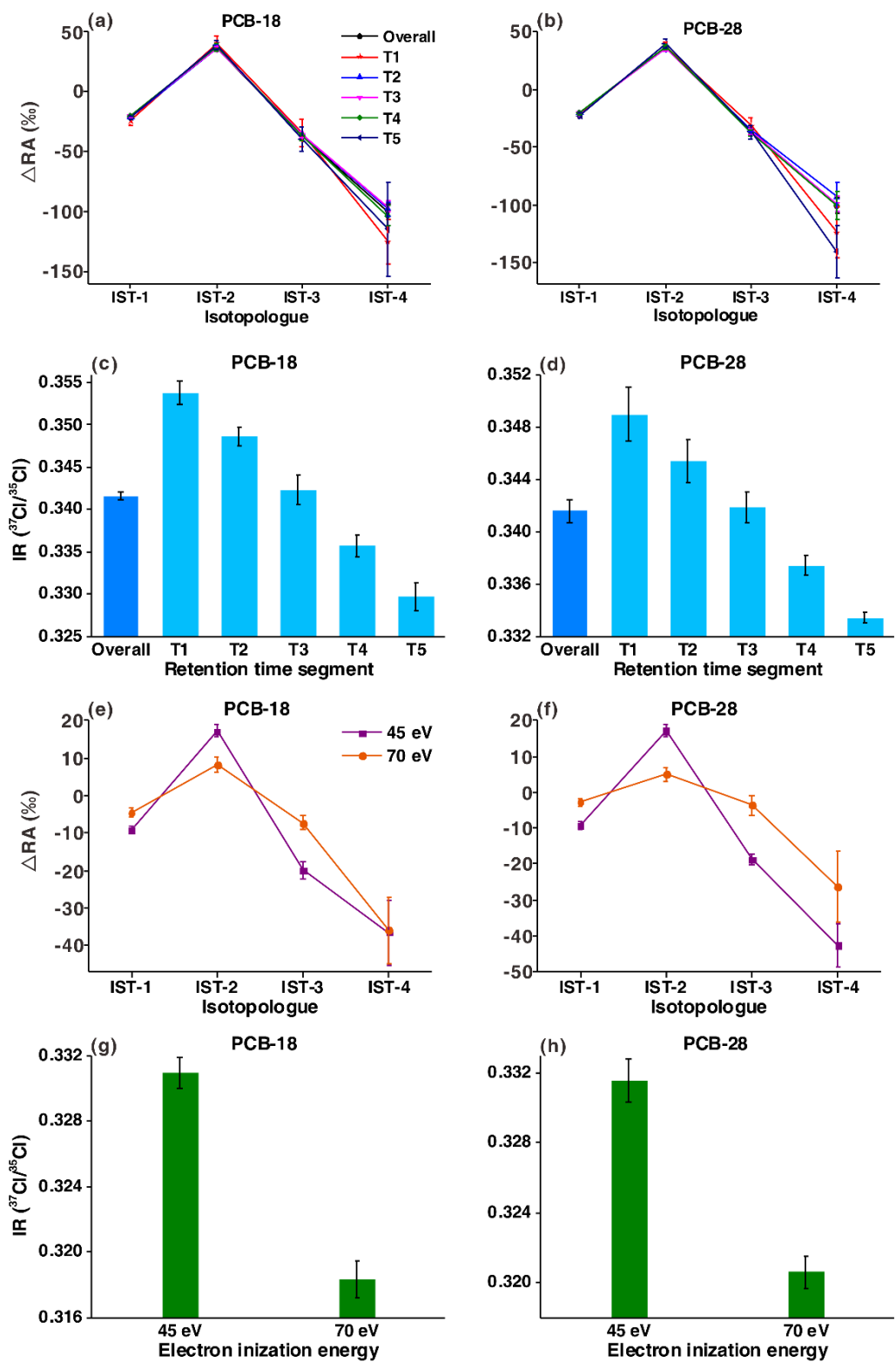

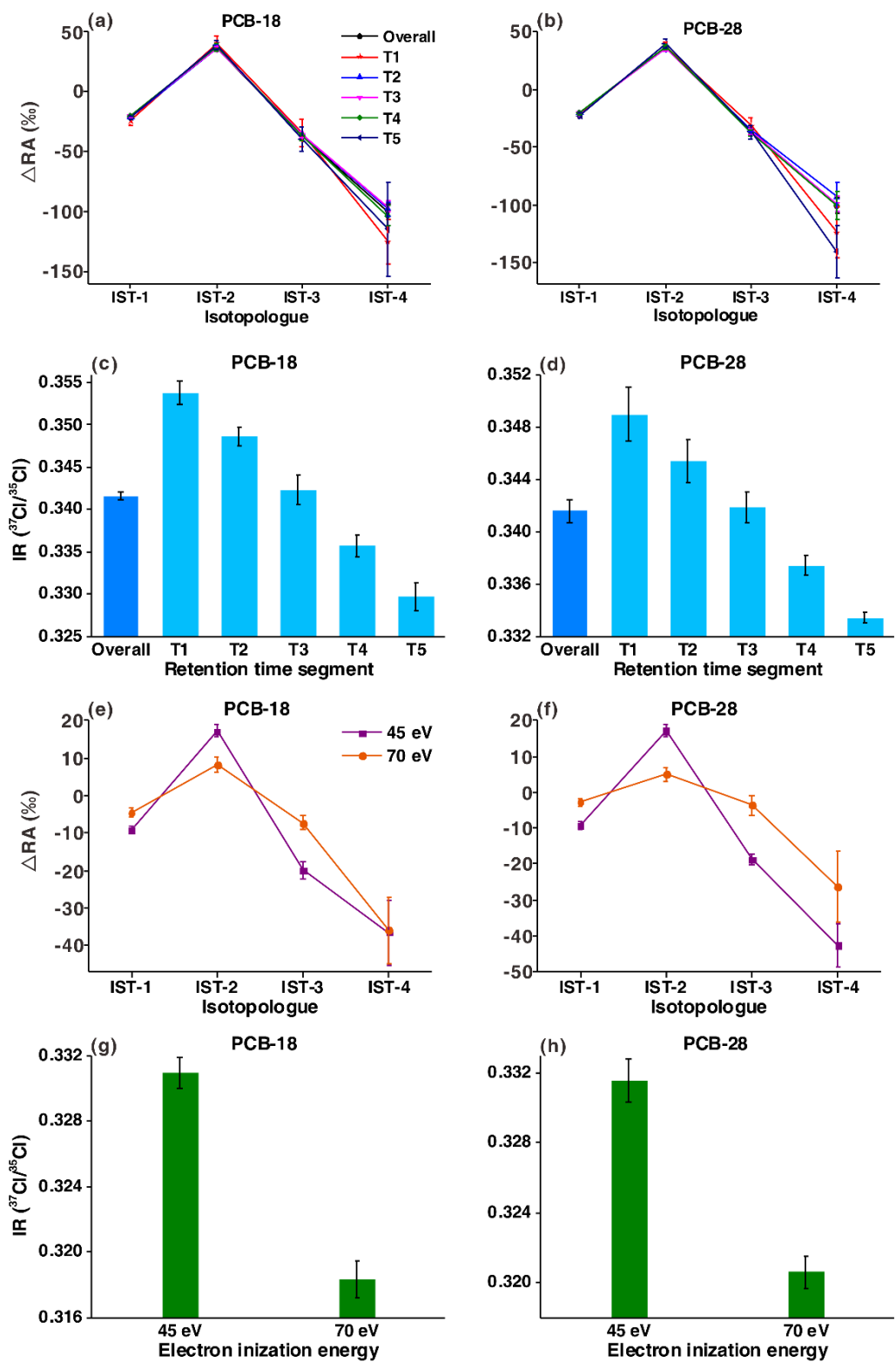

**Figure 7**. ΔRA and isotope ratios of PCB-18 and PCB-28 in different retention-time segments (T1-T5) and at different EI energies with the detection of GC-HRMS. **(a)**: ΔRA of PCB-18 in T1-T5 and in overall retention-time range, **(b)**: ΔRA of PCB-28 in T1-T5 and in overall retention-time range, **(c)**: isotope ratios of PCB-18 in T1-T5 and in overall retention-time range,

**(d)**: isotope ratios of PCB-28 in T1-T5 and in overall retention-time range, **(e)**: ΔRA of PCB-18 at EI energies of 45 eV and 70 eV, **(f)**: ΔRA of PCB-28 at EI energies of 45 eV and 70 eV, **(g)**: isotope ratios of PCB-18 at EI energies of 45 eV and 70 eV, **(h)**: isotope ratios of PCB-28 at EI energies of 45 eV and 70 eV; the standard solutions were at 1.0 μg/mL; the injection replicates were five.

**Implications to source identification.**

Using isotope ratios only may not be enough to absolutely discern sources of organochlorines, because it is possible that the bulk isotope ratios of a compound from different sources are equivalent but the isotopologue distributions are different. To the contrary, it is in principle impossible that the isotopologue distributions are similar while the bulk isotope ratios of a compound stemming from different sources are different. Nevertheless, the ΔRA obtained in this study cannot absolutely indicate the differences or similarities of isotopologue distributions. Different ΔRA patterns definitely indicate different isotopologue distributions, but consistent ΔRA patterns may not point to identical isotopologue distributions. For instance, if the isotopologues of an organochlorine from different sources conform to binomial distribution, then all the ΔRA are equal to zero, no matter what the isotopologue distributions and isotope ratios are. Fortunately, this scenario cannot arise due to that the chlorine isotopologues of detected ions of organochlorines on GC-EI-MS by no means comply with binomial distribution[38]. In another case, as aforementioned, ΔRA may be changeless during the isotope fractionation caused by physical changes, demonstrating that different groups of an organochlorine showing consistent ΔRA patterns may have different bulk isotope ratios and isotopologue distributions. As a consequence, ΔRA and bulk isotope ratios should be associatively implemented for source identification. Identical ΔRA patterns and equal isotope ratios confidently indicate the same source of a target compound in different samples. Identical ΔRA patterns and unequal isotope ratios may indicate that different groups of a target compound in different samples are from the same source but have underwent different physical changes. Inconsistent ΔRA patterns and

4different isotope ratios may demonstrate the different sources of a target compound in different samples, or different chemical-reaction types and/or extents experienced by a target compound in different samples from the same source. ΔRA patterns are highly compound-specific fingerprints of organochlorines from different sources or experiencing specific chemical reaction processes, and capable of providing more credible results in source identification comparing with isotope ratios. As a result, ΔRA patterns in conjunction with isotope ratios can significantly improve validity of the outcomes in source identification for organochlorines.

**Implications to source apportionment.**

Bulk chlorine isotope ratios have been applied to the source apportionment for some organochlorines.[43] Unfortunately, the bulk isotope ratios can only be practicable for the source apportionment of organochlorines from no more than two sources in principle. Nevertheless, the relative abundances of chlorine isotopologues can theoretically be applied to the source apportionment of the organochlorines from more sources. Technically, an organochlorine having $n$ chlorine isotopologues with known relative abundances can be used to quantify the proportions of this compound from $m$ ($m \leq n$) sources.

We hypothesize a scenario that an organochlorine with $n$ isotopologues in a sample stem from $m$ sources, which is equivalent to the scenario that a group of an organochlorine with $n$ isotopologues is mixed with $m$ independent groups of the organochlorine. The following system of equations can be obtained:

$$\begin{pmatrix} a_{11} & \cdots & a_{1j} & \cdots & a_{1m} \\ \vdots & & & & \vdots \\ a_{i1} & \cdots & a_{ij} & \cdots & a_{im} \\ \vdots & & & & \vdots \\ a_{n1} & \cdots & a_{nj} & \cdots & a_{nm} \end{pmatrix} \times \begin{pmatrix} P_1 \\ \vdots \\ P_i \\ \vdots \\ P_m \end{pmatrix} = \begin{pmatrix} A_1 \\ \vdots \\ A_i \\ \vdots \\ A_m \end{pmatrix} \quad (m \leq n) \quad (1)$$

Page 21

where $P_j$ is the proportion of the mixing group j; $a_{ij}$ is the relative abundance of the isotopologue i of the organochlorine in the mixing group j; $A_i$ is the relative abundance of the isotopologue i of the organochlorine in the mixed group. As $a_{ij}$ and $A_i$ can be measured by GC-DFS-HRMS, then the portion $P_j$ can be calculated with eq 1. Taking PCB-18 and PCB-28 for example, we regard the molecules of a PCB in an overall chromatographic peak (the mixed group) as the combination of those from four retention-time segments of the peak (the mixing groups). Then, we have

$$\begin{pmatrix} a_{11} & a_{12} & a_{13} & a_{14} \\ a_{21} & a_{22} & a_{23} & a_{24} \\ a_{31} & a_{32} & a_{33} & a_{34} \\ a_{41} & a_{42} & a_{43} & a_{44} \end{pmatrix} \times \begin{pmatrix} P_1 \\ P_2 \\ P_3 \\ P_4 \end{pmatrix} = \begin{pmatrix} A_1 \\ A_2 \\ A_3 \\ A_4 \end{pmatrix} \qquad (2)$$

The proportions of a PCB from the four retention-time segments ($P_1$-$P_4$) thus can be calculated by solving eq 2. As illustrated in Figure 8, the calculated proportions agree very well with the measured proportions based on the MS signal intensities of the two PCBs, indicating the promising competence of $RA_{mea}$ of chlorine isotopologues in source apportionment for chlorinated organic compounds.

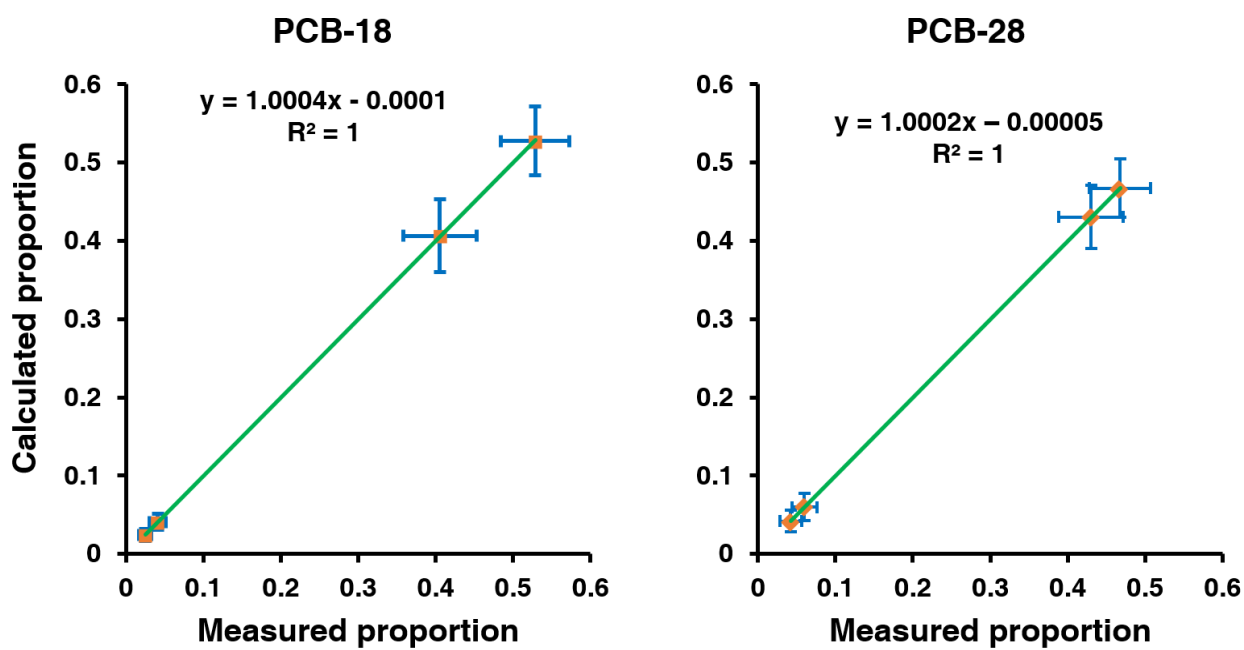

**Figure 8**. Calculated proportions based on the RA$_{mea}$ of chlorine isotopologues and the measured proportions based on the detected MS signal intensities of PCB-18 and PCB-28 derived from four retention-time segments with the detection of GC-HRMS. Error bars represent the standard deviations (0.8%-4.7%, 1σ, n=5). Calculated proportions were obtained through solving eq 2 with Matlab 2016a.

## CONCLUSIONS

This study developed and validated a method for analyzing compound-specific chlorine isotopologue distribution of organochlorines using GC-DFS-HRMS for source identification and apportionment. Overall chlorine isotopologues of the organochlorine analytes were determined with GC-DFS-HRMS operated in MID mode. $RA_{mea}$, $RA_{sim}$ and $\Delta RA$ were obtained based on the detected MS signal intensities of individual isotopologues. The method was partially validated in the aspects of precision, injection-amount dependency and temporal drifts. The $RA_{mea}$ and $\Delta RA$ of the method showed high precision. The $\Delta RA$ and $\Delta RA$ patterns were statistically distinguishable for the PCE and TCE standards from different manufacturers. The $\Delta RA$ did not show observable injection-amount dependency and evident temporal drift with injection times and analysis batches, while presented statistically significant differences on different types of MS and on the same MS with different EI energies. We have successfully applied this method to an analogous scenario of source identification and apportionment of two PCBs that presented chlorine isotope fractionation on the GC-HRMS system. The results showed that the $\Delta RA$ in association with isotope ratios enable source identification for organochlorines with high validity. In addition, the $\Delta RA$ were found to be consistent during the isotope fractionation caused by physical processes while change with chemical reactions. This demonstrates that the $\Delta RA$ can also be applied to differentiating whether isotope fractionation is caused by physical changes or chemical reactions, and further to inferring the chemical reactions by which organochlorines of interest have experienced. Moreover, the $RA_{mea}$ can be used in source apportionment for organochlorines from more sources with more reliable results in comparison with the methods using isotope ratios only. This study provides a new way to fingerprinting analysis of compound-specific chlorine isotopologue distribution of chlorinated organic compounds, and proposes innovative insights into high efficient and reliable source delineation and apportionment for organochlorines.

## ASSOCIATED CONTENT

The Supplementary Information is available free of charge on the website at http://pending.

## AUTHOR INFORMATION

**Corresponding author**

*Email: CaimingTang@gig.ac.cn (C. Tang).

## ACKNOWLEDGEMENTS

We are grateful for Mr. Deyun Liu from Guangzhou Quality Supervision and Testing Institute, China for his assistance in the GC-qMS analysis. This work was partially supported by the National Natural Science Foundation of China (Grant No. 41603092).

**Table 1**. Measured relative abundances (RA$_{mea}$), simulated relative abundances (RA$_{sim}$), relative-abundance variations (ΔRA) and precisions (standard deviations) of chlorine isotopologues of perchlorethylene (PCE) and trichloroethylene (TCE) from different manufacturers in different analysis batches with the detection of GC-HRMS.

| PCE from manufactuers-1 | | | Batch-I | | | | | | Batch-II | | | | | |
|---|---|---|---|---|---|---|---|---|---|---|---|---|---|---|
| Isotopologue | Isotopologue formula | $m/z$ | RA$_{mea}$ (mean) | SD (1σ, n=5, ‰) | RA$_{sim}$ (mean) | SD (1σ, n=5, ‰) | ΔRA (mean, ‰) | SD (1σ, n=5, ‰) | RA$_{mea}$ (mean) | SD (1σ, n=6, ‰) | RA$_{simulated}$ (mean) | SD (1σ, n=6, ‰) | ΔRA (mean, ‰) | SD (1σ, n=6, ‰) |
| IST-1 | $C_2{}^{35}Cl_4$ | 163.87486 | 0.32588 | 0.69 | 0.33088 | 0.39 | -15.11 | 1.17 | 0.32578 | 0.20 | 0.33091 | 0.21 | -15.49 | 0.65 |
| IST-2 | $C_2{}^{35}Cl_3{}^{37}Cl$ | 165.87191 | 0.42965 | 0.63 | 0.42155 | 0.02 | 19.22 | 1.47 | 0.42995 | 0.36 | 0.42155 | 0.01 | 19.93 | 0.87 |
| IST-3 | $C_2{}^{35}Cl_2{}^{37}Cl_2$ | 167.86896 | 0.19993 | 0.33 | 0.20140 | 0.26 | -7.29 | 1.30 | 0.19968 | 0.39 | 0.20138 | 0.13 | -8.46 | 1.55 |
| IST-4 | $C_2{}^{35}Cl{}^{37}Cl_3$ | 169.86601 | 0.04140 | 0.18 | 0.04277 | 0.11 | -31.94 | 4.03 | 0.04147 | 0.18 | 0.04276 | 0.06 | -30.19 | 4.45 |
| IST-5 | $C_2{}^{37}Cl_4$ | 171.86306 | 0.00314 | 0.02 | 0.00341 | 0.01 | -78.17 | 7.61 | 0.00312 | 0.06 | 0.00340 | 0.01 | -82.53 | 16.91 |
| PCE from manufactuers-2 | | | Batch-I | | | | | | Batch-II | | | | | |
| IST-1 | $C_2{}^{35}Cl_4$ | 163.87486 | 0.32505 | 0.22 | 0.32743 | 0.21 | -7.25 | 0.44 | 0.32584 | 0.39 | 0.32818 | 0.29 | -7.13 | 0.48 |
| IST-2 | $C_2{}^{35}Cl_3{}^{37}Cl$ | 165.87191 | 0.42484 | 0.20 | 0.42169 | 0.01 | 7.49 | 0.48 | 0.42482 | 0.21 | 0.42166 | 0.01 | 7.49 | 0.48 |
| IST-3 | $C_2{}^{35}Cl_2{}^{37}Cl_2$ | 167.86896 | 0.20424 | 0.27 | 0.20365 | 0.14 | 2.87 | 0.84 | 0.20366 | 0.21 | 0.20316 | 0.19 | 2.45 | 0.74 |
| IST-4 | $C_2{}^{35}Cl{}^{37}Cl_3$ | 169.86601 | 0.04257 | 0.14 | 0.04371 | 0.06 | -26.09 | 2.89 | 0.04241 | 0.10 | 0.04351 | 0.08 | -25.23 | 2.20 |
| IST-5 | $C_2{}^{37}Cl_4$ | 171.86306 | 0.00329 | 0.02 | 0.00352 | 0.01 | -64.28 | 5.94 | 0.00328 | 0.04 | 0.00349 | 0.01 | -61.37 | 10.37 |
| TCE from manufactuers-1 | | | Batch-I | | | | | | Batch-II | | | | | |
| IST-1 | $C_2H{}^{35}Cl_3$ | 129.91383 | 0.43872 | 0.57 | 0.44007 | 0.54 | -3.06 | 0.33 | 0.43584 | 0.52 | 0.43718 | 0.51 | -3.07 | 0.33 |
| IST-2 | $C_2H{}^{35}Cl_2{}^{37}Cl$ | 131.91088 | 0.41784 | 0.32 | 0.41547 | 0.20 | 5.70 | 0.54 | 0.41886 | 0.33 | 0.41653 | 0.19 | 5.60 | 0.68 |
| IST-3 | $C_2H{}^{35}Cl{}^{37}Cl_2$ | 133.90793 | 0.13005 | 0.27 | 0.13075 | 0.28 | -5.35 | 0.26 | 0.13164 | 0.30 | 0.13229 | 0.27 | -4.86 | 1.15 |
| IST-4 | $C_2H{}^{37}Cl_3$ | 135.90498 | 0.01339 | 0.08 | 0.01372 | 0.05 | -23.58 | 4.89 | 0.01366 | 0.08 | 0.01400 | 0.05 | -24.90 | 2.96 |
| TCE from manufactuers-2 | | | Batch-I | | | | | | Batch-II | | | | | |
| IST-1 | $C_2H{}^{35}Cl_3$ | 129.91383 | 0.43794 | 0.46 | 0.43797 | 0.50 | -0.08 | 0.32 | 0.43475 | 0.46 | 0.43473 | 0.43 | 0.05 | 0.26 |
| IST-2 | $C_2H{}^{35}Cl_2{}^{37}Cl$ | 131.91088 | 0.41630 | 0.25 | 0.41624 | 0.18 | 0.13 | 0.61 | 0.41729 | 0.26 | 0.41742 | 0.15 | -0.32 | 0.47 |
| IST-3 | $C_2H{}^{35}Cl{}^{37}Cl_2$ | 133.90793 | 0.13185 | 0.32 | 0.13186 | 0.27 | -0.07 | 0.74 | 0.13379 | 0.23 | 0.13360 | 0.23 | 1.46 | 0.48 |
| IST-4 | $C_2H{}^{37}Cl_3$ | 135.90498 | 0.01391 | 0.07 | 0.01392 | 0.05 | -0.84 | 2.60 | 0.01417 | 0.06 | 0.01425 | 0.04 | -6.00 | 2.74 |

Note: the standard solutions were at 1.0 μg/mL; $m/z$: mass to charge ratio, SD: standard deviation, IST: isotopologue.

## Table of Contents

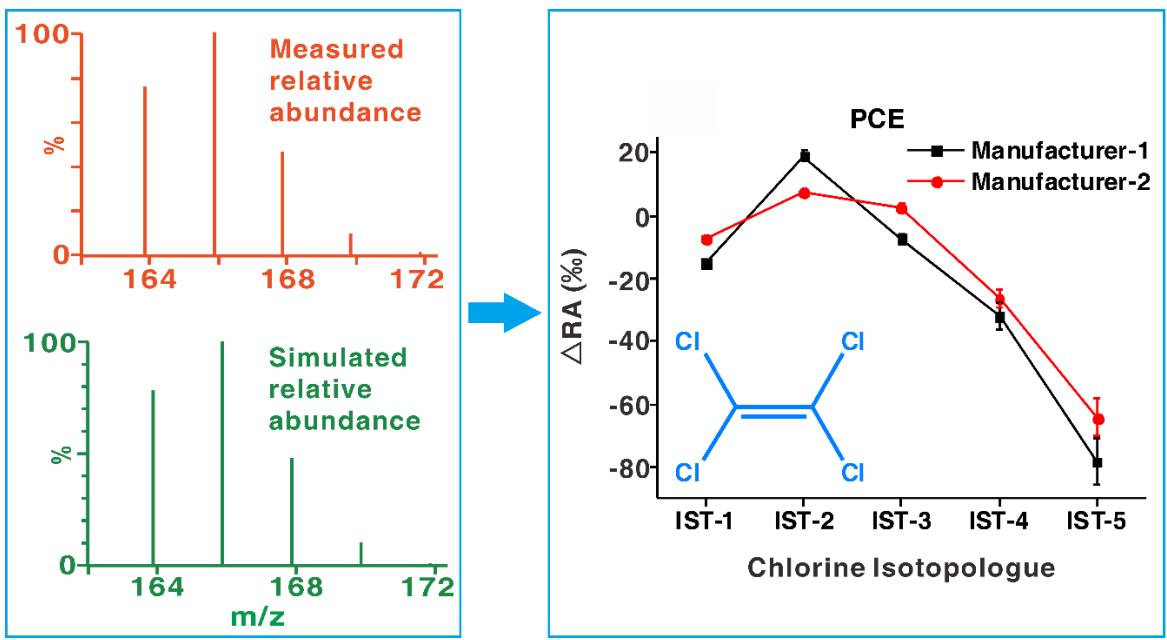